\documentclass[journal=jacsat,manuscript=article]{achemso}

\usepackage[version=3]{mhchem} 
\usepackage{amsfonts}

\author{Felix Eder}
\affiliation[Unige]
{University of Geneva, Department of Quantum Matter Physics, 24 Quai Ernest-Ansermet, CH-1211 Geneva, Switzerland}
\author{Catherine Witteveen}
\affiliation[Unige]
{University of Geneva, Department of Quantum Matter Physics, 24 Quai Ernest-Ansermet, CH-1211 Geneva, Switzerland}
\author{Enrico Giannini}
\affiliation[Unige]
{University of Geneva, Department of Quantum Matter Physics, 24 Quai Ernest-Ansermet, CH-1211 Geneva, Switzerland}
\author{Fabian O. von Rohr}
\affiliation[Unige]
{University of Geneva, Department of Quantum Matter Physics, 24 Quai Ernest-Ansermet, CH-1211 Geneva, Switzerland}
\email{correspondance}

\title[]
  {Stoichiometry and Phase Control in K\textsubscript{1--\textit{x}}CrSe\textsubscript{2} via Self-Flux Synthesis}

\begin{document}




\begin{abstract}
Layered delafossite-type magnetic materials, such as \ce{KCrSe2}, are promising platforms for studying magnetic systems and potential frustration on triangular lattices. Synthesis, structure-type control, and off-stoichiometries remain major challenges in the investigation of these delafossite-type magnets. Starting from the same self-flux composition (K:Cr:Se = 8:1:8), we isolated three distinct \ce{K_{1--\textit{x}}CrSe2} phases with \textit{x} = 0, 0.13--0.17, and 0.32--0.35, each adopting a different structure type depending on the quenching temperature applied. The phase evolution indicates a sequence of transformations during synthesis between compounds with varying degrees of potassium deficiency. Building on these insights into phase stability and crystal growth, we successfully grew single crystals of full-stoichiometric \ce{KCrSe2} -- enabling direction-dependent magnetization measurements. These measurements reveal a pronounced field dependence of the N\'{e}el temperature at low external fields, as well as a weak metamagnetic transition. Our findings demonstrate that even a simple parameter -- such as quenching temperature -- can be used to control stoichiometry, direct phase formation, and ultimately tune the magnetic properties of delafossite-type materials.

\end{abstract}

\section{Introduction}

Layered magnetic materials with intrinsically two-dimensional structural motifs offer fertile ground for tunable emergent phenomena, including frustrated magnetism, and field-dependent ordering.\cite{Burch2018,Gibertini2019,chamorro2020chemistry,Wolfspintronics2001} Among these, compounds featuring triangular lattices of magnetic cations are compelling, as their geometry promotes magnetic frustration and complex spin interactions. Delafossite-type structures exemplify this class, comprising stacked layers of edge-sharing \ce{\textit{TMX}6} octahedra (\textit{TM} = transition metal, \textit{X} = chalcogen), separated by cation sheets that limit interlayer coupling and enhance the two-dimensionality of their properties.\cite{moll2016evidence,zhang2024crystal}

Within this broader class, the family of chromium-based delafossite-type compounds \ce{\textit{A}Cr\textit{X}2} (\textit{A} = monovalent cation, \textit{X} = S, Se, or Te) has attracted particular attention due to its chemical tunability and varied magnetic ground states.\cite{nocerino2023competition,Kobayashi2019,song2019soft} Their crystal structures feature triangular \ce{Cr\textit{X}2} layers separated by intercalated cation sheets, yet small changes in composition can strongly affect coordination environments, interlayer distances, and ultimately magnetic interactions.\cite{Kobayashi2019} While in the field of the structurally closely related materials like \ce{Na_{\textit{x}}CoO2} or \ce{Li_{\textit{x}}CoO2} structure-property relationships have been studied extensively for various \textit{x} contents for battery applications as well as collective quantum properties\cite{Sakurai2004,Hertz2008}, off-stoichiometry in the isostructural selenide phases is gaining attention recently, exemplified by the structural richness of the \ce{Sn_{\textit{x}}TaSe2} system.\cite{Bierman2025}

In this context, potassium chromium diselenide (\ce{KCrSe2}) and its potassium-deficient variants offer a case study, in which changes in stoichiometry and synthesis conditions stabilize distinct structural and magnetic ground states. Early reports identified the full-stoichiometric \ce{KCrSe2} as rhombohedral (\textit{R$\bar{3}$m}). This phase acts commonly as precursor for oxidative deintercalation routes to the van der Waals magnet \ce{CrSe2}, which can only be accessed through kinetically controlled synthesis methods.\cite{VanBruggen1980,Song2021} Magnetic measurements of \ce{KCrSe2} have found A-type antiferromagnetic order with reported Néel temperatures ranging from \textit{T}\textsubscript{N} = 40 K to 86 K.\cite{Wiegers1980,Fang_1996,Song2021} In addition to the stoichiometric compound, also potassium-deficient phases as \ce{K_{1--\textit{x}}CrSe2} have been reported. These include a rhombohedral phase with partially occupied \ce{K} sites and expanded interlayer spacing \ce{K_{0.6--0.8}CrSe2}, and a incompletely resolved orthorhombic variant reported near \ce{K_{0.9}CrSe2}.\cite{Nikiforow1991,Wiegers1980} The latter might potentially correspond to the latest addition to the \ce{K_{1--\textit{x}}CrSe2} phase system: we have recently reported the growth and structural refinement of an incommensurately modulated monoclinic phase, \ce{K_{0.83–0.87}CrSe2}, using K/Se self-flux synthesis followed by hot-centrifugation.\cite{Eder2025} This phase shows an atypical AB layer stacking, rather than the typical ABC stacking observed in delafossite-type materials. Its crystal structure can be described using one incommensurate modulation vector, which is directly linked to its potassium deficiency, resulting in a 3+1 dimensional crystal structure. Its magnetic behavior also differs, with a strongly enhanced AFM order at \textit{T}\textsubscript{N} = 134 K. The synthetic conditions under which these various \ce{K_{1--\textit{x}}CrSe2} phases emerge, and how potassium content, structural distortion, and magnetic behavior are connected, have remained unclear.

Flux synthesis has become an indispensable method in solid-state chemistry not only for accessing high-quality single crystals\cite{canfield2001high,guguchia2017signatures,witteveen2023synthesis}, but likewise as a tool for the discovery of new phases.\cite{kanatzidis2005metal,bugaris2012materials,lefevre2022heavy,lefevre2024new} In particular, self-flux techniques, where at least one of the reactants serves as both a reagent and solvent, allow for milder growth conditions and cleaner separation of products.\cite{Aitken1998,Sturza2014,Martin2004}

In the case of \ce{K_{1--\textit{x}}CrSe2}, we found that a K/Se self-flux favours the crystal growth and hot centrifugation at carefully controlled temperatures enables to tune the crystal stoichiometry and governs the phase selectivity. Building on our prior isolation of an incommensurately modulated phase at $T$ = 750 °C,\cite{Eder2025} we now show that systematic variation of the quenching temperature gives reproducible access to three distinct compounds, including stoichiometric \ce{KCrSe2}, the incommensurately modulated \ce{K_{0.83--0.87}CrSe2} and the vacancy-disordered phase \ce{K_{0.65--0.68}CrSe2}. Our findings further show that alkali/chalcogenide mixed flux synthesis is not merely a growth method, but a synthetic handle for navigating the structural landscape of delafossite-type materials.

\section{Experimental}

\textit{Synthesis:} All phases reported in this article were synthesized from the same combination of starting materials. Reactant and product handling was performed in an Ar-filled glovebox. Potassium (block, Sigma Aldrich, 99\%), chromium (powder, Alfa Aesar, 99.99\%), and selenium (pieces, Alfa Aesar, 99.999\%) were weighed in molar ratios of 8:1:8 (total mass 1.000 g) in an alumina Canfield crucible set\cite{Canfield2016}. For the optimized synthesis of \ce{K_{0.65(3)--0.68(2)}CrSe2}, a reduction of the K-content of the flux down to molar ratios of 5:1:8 was employed.
The reactants were then sealed in quartz ampoules under a partial Ar atmosphere (ca. 300 mbar).
After this, the reaction setup was placed in a muffle furnace and heated to the maximum temperature of 1000 °C at a rate of 60 °C/h. After dwelling at maximum temperature for 24 h, a slow cooling rate of 3 °C/min was employed until the quenching temperature \textit{T} was reached. The products were held at this temperature for several hours before the liquid flux was separated from the crystals by hot-centrifugation. For some samples, an extended holding time \textit{t} at the quenching temperature before centrifugation was kept to highlight the influence of reaction time at a given temperature.

\textit{Analysis:} The elemental composition of the products was checked using energy dispersive X-ray spectroscopy (EDS) on a JEOL JSM-7600F scanning electron microscope (SEM) with an accelerating voltage of 20 kV. 
Powder X-ray diffraction (PXRD) measurements were performed on a Rigaku SmartLabXE diffractometer equipped with a D/teX Ultra 250 detector using Cu-K\textsubscript{$\alpha$} radiation in Debye-Scherrer geometry. Samples were manually ground under a protective atmosphere and filled into glass capillaries with an outer diameter of 0.5 mm.
PXRD data containing \ce{K_{0.83--0.87}CrSe2} as the main phase were instead collected on a
Rigaku Smart-Lab diffractometer equipped with a 9 kW PhotonMax rotating anode Cu-source.
The crystal structures of \ce{KCrSe2}, \ce{K_{0.68(2)}CrSe2} and \ce{K_{0.65(3)}CrSe2} were investigated by single crystal X-ray diffraction (SXRD) on a Rigaku Supernova diffractometer using Mo-K\textsubscript{$\alpha$} radiation at 100 K. For that of \ce{K_{0.83--0.87}CrSe2}, we refer to our recent investigation of its incommensurately modulated crystal structure.\cite{Eder2025} 
Unit-cell indexation, integration, and numerical absorption correction based on crystal faces were performed with CrysalisPro, \cite{CrysAlisPRO2022} structure solution and refinement were done with SHELXT and SHELXL.\cite{sheldrick_shelxt_2015,sheldrick_shelxl_2015}
Magnetic property measurements were performed in a QuantumDesign Dynacool PPMS equipped with a 9 T magnet and a vibrating sample magnetometry (VSM) option. Temperature-dependent measurements were conducted in a temperature range between $T$ = 2 and 300 K in sweep mode at a heating rate of 5 K/min, field-dependent measurements from $\mu_0 H$ = --9 to 9 T in sweep mode at 50 G/s.

\section{Results and discussion}

\subsection{Synthesis and SXRD}

\begin{figure*}
\begin{center}
\includegraphics[width=\textwidth]{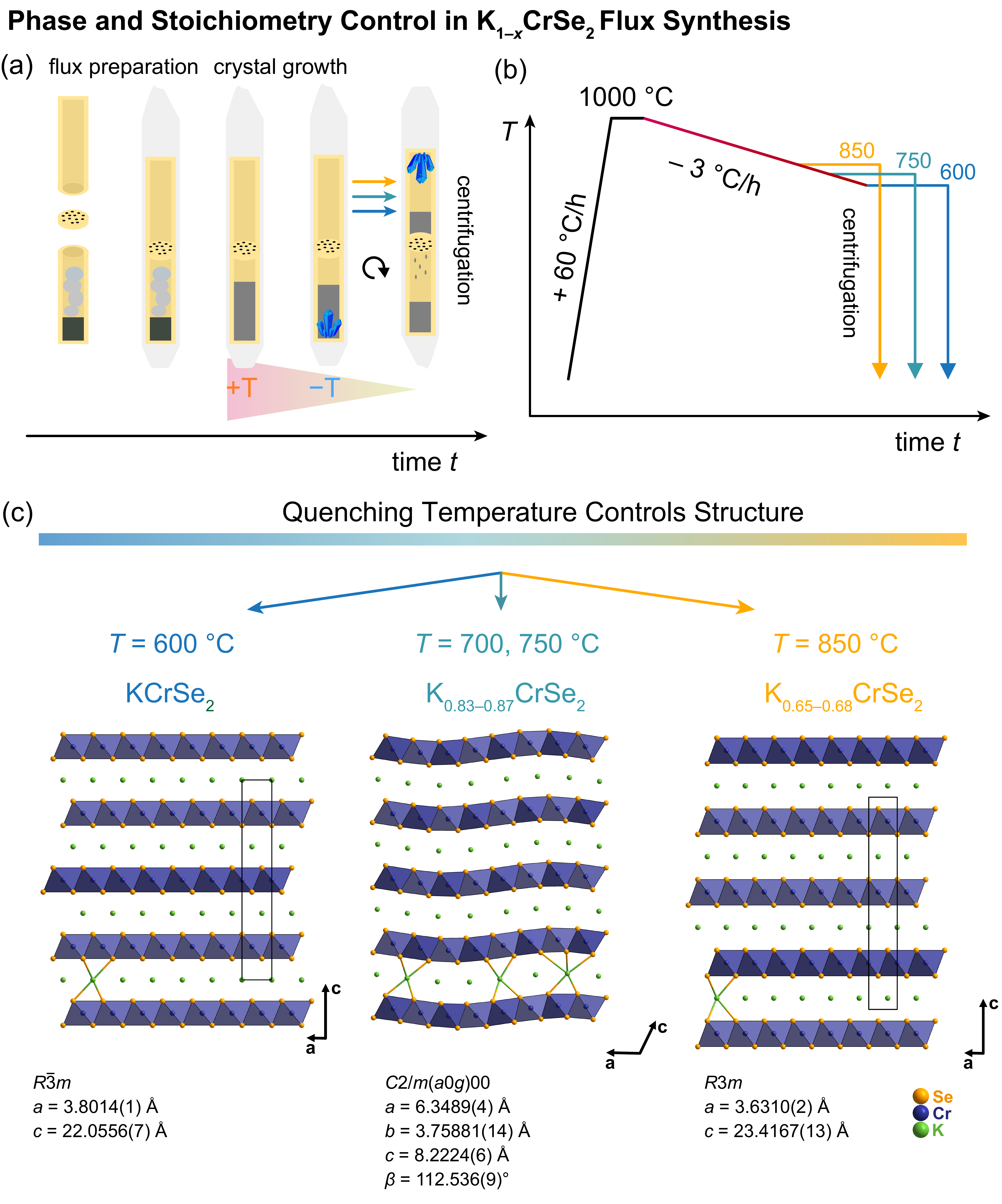}
\caption{Phase and Stoichiometry Control: (a) Scheme of self-flux synthesis consisting of assembly of a Canfield crucible set, sealing the setup in a quartz ampoule, heating up to achieve a homogenous melt, slow cooling to grow large crystals, and quenching during hot-centrifugation. (b) The temperature profile applied for synthesis of various \ce{K_{1--\textit{x}}CrSe2} phases. 
(c) The crystal structures of full stoichiometric \ce{KCrSe2}, incommensurately modulated \ce{K_{0.866}CrSe2} and \ce{K_{0.68(2)}CrSe2} viewed along \textbf{b}. The different coordinations of K cations by Se atoms of the \ce{CrSe2} layers are highlighted at the bottom.}
\label{Fig1}
\end{center}
\end{figure*}

We employed a K/Se self-flux approach using a nominal composition of K:Cr:Se = 8:1:8, following the procedure illustrated in Fig. \ref{Fig1}(a). The reaction mixtures were sealed in Canfield crucible sets, heated to 1000 °C, and then slowly cooled to facilitate crystal growth. Crystals were isolated from the remaining K/Se flux by hot centrifugation at varying temperatures (\textit{T} = 600–-850 °C), as shown in the applied temperature profile (Fig. \ref{Fig1}(b)). We found the quenching temperature to be a decisive parameter for phase formation (Fig. \ref{Fig1}(c)): at a centrifugation temperature of \textit{T} = 600 °C, we obtained phase-pure, stoichiometric \ce{KCrSe2}; at 700--750 °C, the incommensurately modulated phase \ce{K_{0.83--0.87}CrSe2} formed; and at 800--850 °C, crystals with compositions in the range \ce{K_{0.65--0.68}CrSe2} were recovered. Hereby, a reduction of the K-content of the applied flux down to a K:Cr:Se ratio of 5:1:8 proved to significantly increase the \ce{K_{0.65--0.68}CrSe2} content in the reaction product.
All obtained crystals are found to be platelets of silvery metallic luster. While for \ce{KCrSe2}, platelets with lateral dimensions of several mm were obtained, those of \ce{K_{0.65--0.68}CrSe2} clearly failed to reach mm sizes, which would be needed for anisotropic magnetization measurements.

The three phases could be well-identified using SXRD. These three compositionally distinct phases correspond to three different crystal structures: In addition to our detailed report of SXRD data on the incommensurately modulated \ce{K_{0.83--0.87}CrSe2} phases,\cite{Eder2025} we have collected SXRD data for both the stoichiometric \ce{KCrSe2} and the potassium-deficient phases \ce{K_{0.68(2)}CrSe2} and isostructural \ce{K_{0.65(3)}CrSe2} at 100 K (Table \ref{Crystdata}). Single crystals of \ce{K_{0.68(2)}CrSe2} and \ce{K_{0.65(3)}CrSe2} were taken from the same synthetic batch, highlighting the compositional variability and the inhomogeneity of growth conditions during self-flux synthesis. This is in line with the compositional inhomogeneity we had observed for \ce{K_{0.83--0.87}CrSe2}\cite{Eder2025}. There, the K-content was directly linked to the size of the modulation vector \textbf{q}.
In the case of \ce{K_{0.65--0.68}CrSe2}, we expect the stoichiometry region, in which the phase exists, to be even larger, as the structure type has previously been reported for the range \ce{K_{0.6--0.8}CrSe2}.\cite{Wiegers1980}

While the space group of \ce{KCrSe2} was initially determined to be $R\bar{3}m$\cite{Fang_1996}, most recent investigations have chosen its monoclinic subgroup $C2/m$ instead due to small distortions in unit-cell dimensions and interatomic distances.\cite{Song2021,li2023} We have performed SXRD refinements in both space groups and did not observe sufficient evidence to follow the lower-symmetric space group. A more detailed discussion on the space-group choice, also concerning \ce{K_{0.68(2)}CrSe2} and \ce{K_{0.65(3)}CrSe2} can be found in the SI.

The crystal structure of \ce{KCrSe2} is refined accurately in the trigonal $R\bar{3}m$ space group, consistent with earlier reports \cite{Wiegers1980}. 
The crystal structures of \ce{K_{0.68(2)}CrSe2} and \ce{K_{0.65(3)}CrSe2} were refined in $R3m$ with significantly contracted in-plane lattice parameter \textit{a} and expanded interlayer spacing indicated by an increas in \textit{c} -- in comparison with the stoichiometric form. While this phase was first described decades ago from powder diffraction data \cite{Wiegers1980}, our results provide the single-crystal structure determination of this highly K-deficient phase. The crystal structures of the three phases are depicted in Fig. \ref{Fig1}(c). CIF-files of the three structural solutions can be accessed in the supporting information and are available in the CCDC database under the deposition numbers given in Table \ref{Crystdata}.

\begin{table}
	\begin{center}
	\caption{Crystallographic data for single-crystals of \ce{KCrSe2} and \ce{K_{0.68(2)}CrSe2}.}\label{Crystdata}
		\begin{tabular}{llll}
 & \textbf{\ce{KCrSe2}} & \textbf{\ce{K_{0.68(2)}CrSe2}} & \textbf{\ce{K_{0.65(3)}CrSe2}}\\
 \hline
Chemical formula & \ce{KCrSe2}  & \ce{K_{0.68(2)}CrSe2} & \ce{K_{0.65(3)}CrSe2}\\
Mol. mass (g\; mol$^{-1}$) & 249.02 & 236.38 & 235.26\\ 
Cryst. syst. & trigonal & trigonal & trigonal\\
Space group & $R\Bar{3}m$ (166) & $R3m$ (160) & $R3m$ (160)\\
$a$ (\AA) & 3.80140(10) & 3.6310(2) & 3.6052(2)\\ 
$c$ (\AA) & 22.0556(7) & 23.4167(13) & 23.707(3)\\ 
$V$ (\AA$^{3}$) & 267.017(17) & 267.37(3) & 266.85(4)\\ 
$Z$ & 3 & 3 & 3\\
Calculated density (g\; cm$^{-3}$) & 4.494 & 4.404 & 4.392\\
Temperature (K) & 100(2) & 100(2) & 100(2)\\
Diffractometer & \multicolumn{3}{l}{Oxford Diffraction Supernova} \\
Radiation ($\lambda$) & \multicolumn{3}{l}{Mo K$\alpha$ (0.71073 \AA)} \\
Crystal color & grey & grey & grey\\
Crystal description & plate & plate & plate\\
Crystal size (mm$^{3}$) & 0.18 $\times$ 0.13 $\times$ 0.05 & 0.23 $\times$ 0.12 $\times$ 0.07 & 0.16 $\times$ 0.09 $\times$ 0.04\\
Linear absorption coefficient (mm$^{-1}$) & 23.749 & 24.136 & 24.149\\
Scan mode & \multicolumn{3}{l}{$\omega$} \\
$\theta$\textsubscript{min}--$\theta$\textsubscript{max} (°) & 2.771 -- 34.616 & 2.609 -- 34.828 & 2.577 -- 32.598\\
$h$ range & --5 to 5 & --5 to 5 & --5 to 5\\
$k$ range & --5 to 5 & --5 to 5 & --5 to 5\\
$l$ range & --36 to 36 & --34 to 34 & --33 to 35\\
Measured reflections & 6996 & 7192 & 1822 \\
Completeness (\%) & 98.9 & 98.4 & 97.5 \\
Independent reflections & 180 & 361 & 306\\
\textit{R}\textsubscript{int} & 0.1111 & 0.0727 & 0.0526\\
Absorption correction & numerical & numerical & numerical\\
Independent reflections \\ with I $\geq$ 2.0$\sigma$ & 178 & 354 & 286\\
$R1$ (obs / all) (\%) & 1.82 / 1.83 & 3.06 / 3.10 & 3.75 / 4.00\\
$wR2$ (obs / all) (\%) & 4.81 / 4.82 & 7.57 / 7.60 & 9.38 / 9.52\\
$GOF$ & 1.222 & 1.096 & 1.127\\
Refined parameters & 9 & 15 & 15\\
Maximum difference peaks ($e^-$\AA$^{-3}$) & --1.02; +0.97 & --1.61; +1.36 & --0.70; +1.50\\
CCDC Deposition code & 2475918 & 2475919 & 2475917 \\
    \hline
  \end{tabular}
	\end{center}
\end{table}

The lattice parameters of the under-stoichiometric \ce{K_{0.68(2)}CrSe2} and \ce{K_{0.65(3)}CrSe2} compounds are distorted towards smaller \textit{a} and larger \textit{c} values. This follows a general trend when comparing the unit-cell dimensions and interatomic distances of \ce{K_{1--\textit{x}}CrSe2} phases (Table  \ref{details}).
The contraction of the in-plane lattice parameters between \ce{KCrSe2} and the deintercalated end-member \ce{CrSe2} is hereby much stronger than it would be expected by the different ionic radii \cite{shannon1976} of Cr\textsuperscript{III} and Cr\textsuperscript{IV} ($\Delta$ = 0.065 \AA) or the difference in Cr---Se bond lengths ($\Delta$ = 0.10 \AA) in the two structures.

An explanation for this enhanced in-plane contraction of the Cr---Cr distances lies in the drastically larger ionic radius of the K\textsuperscript{+} cations compared to Cr\textsuperscript{III}. For a coordination number (CN) of 6, which both coordination centers exhibit, this discrepancy amounts to 1.52 vs. 0.755 \AA\cite{shannon1976}. This mismatch leads on one hand, to the pronounced elongation of the trigonal antiprismatic \ce{KSe6} octahedra along \textbf{c}, and on the other hand, the \ce{CrSe2} layers widen up as well to accommodate the sterically dominant alkali metal cations. This effect can be traced from the in-plane lattice parameters \textit{a} for \ce{\textit{A}CrSe2} phases, which are $\sim$3.653 \AA \cite{Kobayashi2019}, 
$\sim$3.732 \AA\cite{engelsman1973} and $\sim$3.801 \AA\, for \textit{A} = Li, Na and K, respectively. With decreasing K-content, the crystal structure can release some of the structural strain by the undulation of \ce{CrSe2} layers in modulated \ce{K_{0.83--0.87}CrSe2}. In \ce{K_{0.65--0.68}CrSe2}, the decreased K:Cr ratio shifts the structural proportions towards the shorter in-plane distances preferred by the Cr atoms.
As the Cr---Se distances remain relatively unchanged, this pushes the Se atoms further away from the layer of triangularly arranged Cr atoms and increases the interlayer distance and therefore the out-of-plane lattice parameter \textit{c}. In this process, the Cr---Se---Cr angles are decreasing.
A much smaller change of the in-plane lattice parameters of \textit{ATM}Se\textsubscript{2} upon deintercalation appears, if the ionic radii and resulting \textit{TM}---Se and \textit{A}---Se bond lengths are in closer agreement. This can for example be observed in \ce{LiZrSe2}
(\textit{a} $\approx$ 3.729 \AA) and the deintercalated \ce{ZrSe2} (\textit{a} $\approx$ 3.771 \AA)\cite{Dahn1985} or for \ce{LiTiSe2} (\textit{a} $\approx$ 3.633 \AA) and \ce{TiSe2} (\textit{a} $\approx$ 3.541 \AA).\cite{patel1983}

\begin{table}
  \caption{Comparison of interatomic distances in \ce{K_{1--\textit{x}}CrSe2} phases including literature references. The first four datasets were recorded at 100 K, while for \ce{CrSe2}, the room-temperature modification was chosen for simplicity. Despite at times higher measurement accuracy, only the first four digits are given for clarity. The appearance of multiple distance values is due to distortion of the coordination octahedra due to symmetry reduction (see Fig. S3)}
  \label{details}
  \begin{tabular}{llllll}
 & \ce{KCrSe2} &  \ce{K_{0.8662(2)}CrSe2}\cite{Eder2025} & \ce{K_{0.68(2)}CrSe2} & \ce{K_{0.65(3)}CrSe2} & \ce{CrSe2}\cite{Kobayashi2019} \\
    \hline
\textit{d} (Cr---Cr) (\AA) & 3.801 & 3.685--3.725, 3.759 & 3.631 & 3.605 & 3.393 \\
\textit{d} (Cr---Se) (\AA) & 2.570 & 2.504--2.582 & 2.507, 2.552 & 2.492, 2.554 & 2.470 \\
$\phi$ (Cr---Se---Cr) (°) & 95.38 & 91.40--96.12 & 90.71, 92.82 & 89.82, 92.63 & 86.77 \\
\textit{d\textsubscript{layer}} (\AA) & 7.352 & 7.594 & 7.806 & 7.902 & 5.915* \\
K coordination & antiprismatic & mixed & prismatic & prismatic & -- \\
Layer stacking & \textit{ABC} & \textit{AB} & \textit{ABC} & \textit{ABC} & \textit{AAA} \\
    \hline
  \end{tabular}
\end{table}

The other notable difference in the various crystal structures of \ce{K_{1--\textit{x}}CrSe2} concerns the coordination of the K cations by the Se atoms of the \ce{CrSe2} layers (Fig. \ref{Fig1}(c)). In \ce{KCrSe2}, they are coordinated in a trigonal antiprismatic geometry, which changes to trigonal prismatic in \ce{K_{0.65--0.68}CrSe2}. In between, for \ce{K_{0.83--0.87}CrSe2}, a wide variety of coordination environments is realized, including CNs of 6, 5+1 and 5+2, and trigonal prismatic, antiprismatic geometries, as well as intermediate surroundings.

In \ce{K_{0.68(2)}CrSe2} and \ce{K_{0.65(3)}CrSe2}, the under-stoichiometry of the K cations is reflected by a reduced site occupation factor of the K1 position. While in \ce{K_{0.83--0.87}CrSe2}, the under-stoichiometry leads to a larger spacing of the K cations from each other and acts as the driving force of the incommensurate modulation, in \ce{K_{0.68(2)}CrSe2}, we have a model of disordered vacancies instead. For some crystals of \ce{K_{0.65(3)--0.68(2)}CrSe2}, we observed weak signs of satellite reflections as well. However, their intensities were not sufficient to give a qualitative description of a superstructure, and even less a quantitative structural refinement.

\subsection{Phase purity}

\begin{figure*}
\begin{center}
\includegraphics[width=0.4\linewidth]{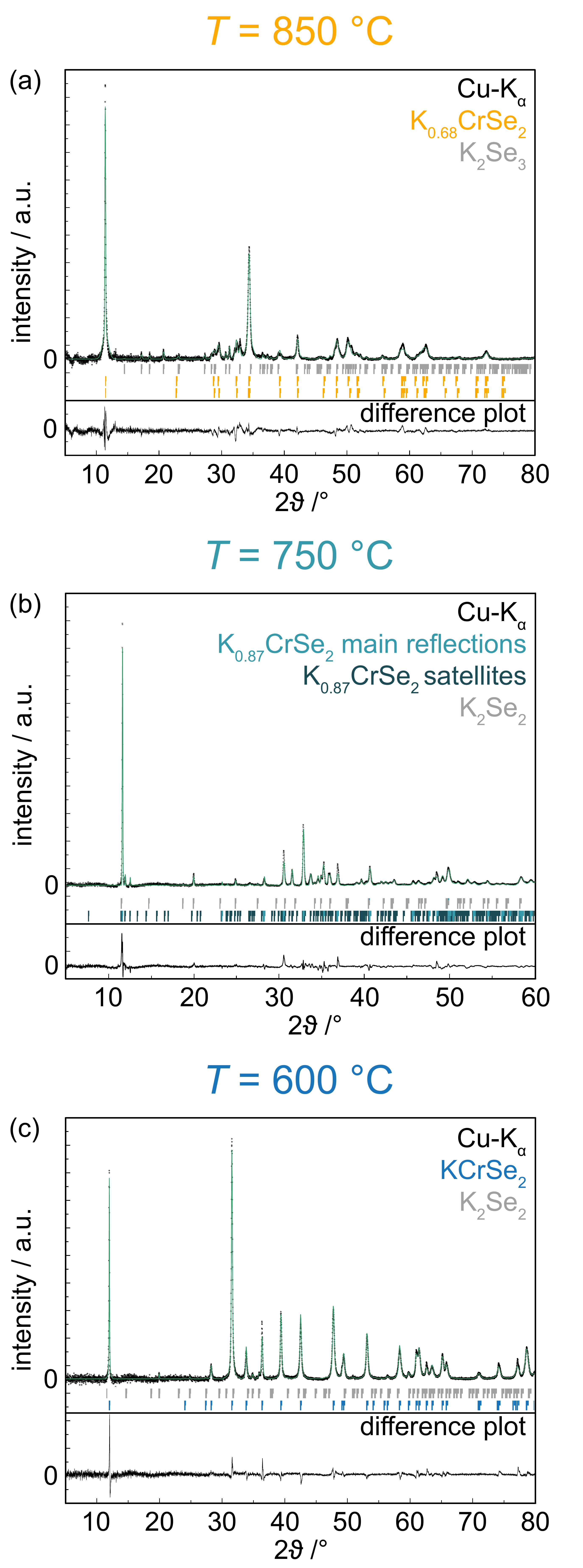}
\caption{PXRD patterns acquired from samples quenched at (a) 850 °C (b) 750 °C (reproduced from [\citenum{Eder2025}]. Copyright 2025 American Chemical Society.), and (c) 600 °C. Background was subtracted after Rietveld refinement for clarity. In all cases, a minor impurity from residual K/Se flux is visible. The three patterns differ markedly in both peak positions and intensities, enabling clear distinction between the resulting phases. These differences reflect changes in symmetry and lattice parameters associated with varying potassium content.}
\label{Fig2}
\end{center}
\end{figure*}

PXRD was used to assess the phase purity across the synthesis conditions and to complement the structural analysis by SXRD. In Fig. \ref{Fig2}, we show the powder diffraction patterns of samples quenched at 850 °C (K:Cr:Se = 5:1:8; (a)), at 750 °C (K:Cr:Se = 8:1:8; (b)), and at 600 °C (K:Cr:Se = 8:1:8; (c)). The three patterns differ markedly in both peak positions and intensities, reflecting the distinct lattice parameters and symmetry changes among the three phases. Thus, PXRD offers a reliable way to distinguish between the three phases.

In all three cases, a small impurity originating from residual K/Se flux is visible, which was not removed completely during centrifugation. For samples quenched from 600 °C, the PXRD pattern confirms the formation of stoichiometric \ce{KCrSe2} with slightly larger lattice parameters than in the SXRD measurements (\textit{a} = 3.80583(5) \AA, \textit{c} = 22.2222(4) \AA\, compared to \textit{a} = 3.80140(10) \AA, \textit{c} = 22.0556(7) \AA). This difference is due to thermal expansion between the different temperatures of data acquisition (100 K for SXRD, and RT for PXRD). 
Reactions quenched at 800 °C yielded multiphase samples composed of \ce{K_{0.68}CrSe2} and incommensurately modulated \ce{K_{0.83--0.87}CrSe2}, as evidenced by overlapping reflections from both phases. Both a reduction of the K-content of the flux and an increase in the quenching temperature managed to decrease the content of the \ce{K_{0.83--0.87}CrSe2} phase, resulting in the pattern displayed in Fig. \ref{Fig2}(a) absent of the modulated phase.
A satisfying Rietveld refinement of the sample quenched from 850 °C (\ce{K_{0.65--0.68}CrSe2}) is complicated by the apparent structural inhomogeneity of the sample, which has already been highlighted by the different lattice parameters for crystals with slightly varying K-contents in SXRD and some deviation in the K-content determined from SEM-EDS (see SI). For the final model, we have included two (\ce{K_{0.65--0.68}CrSe2}) phases to partially account for this issue.

\subsection{Influence of annealing time on the formation of \ce{KCrSe2}}

\begin{figure*}
\begin{center}
\includegraphics[width=16.5cm]{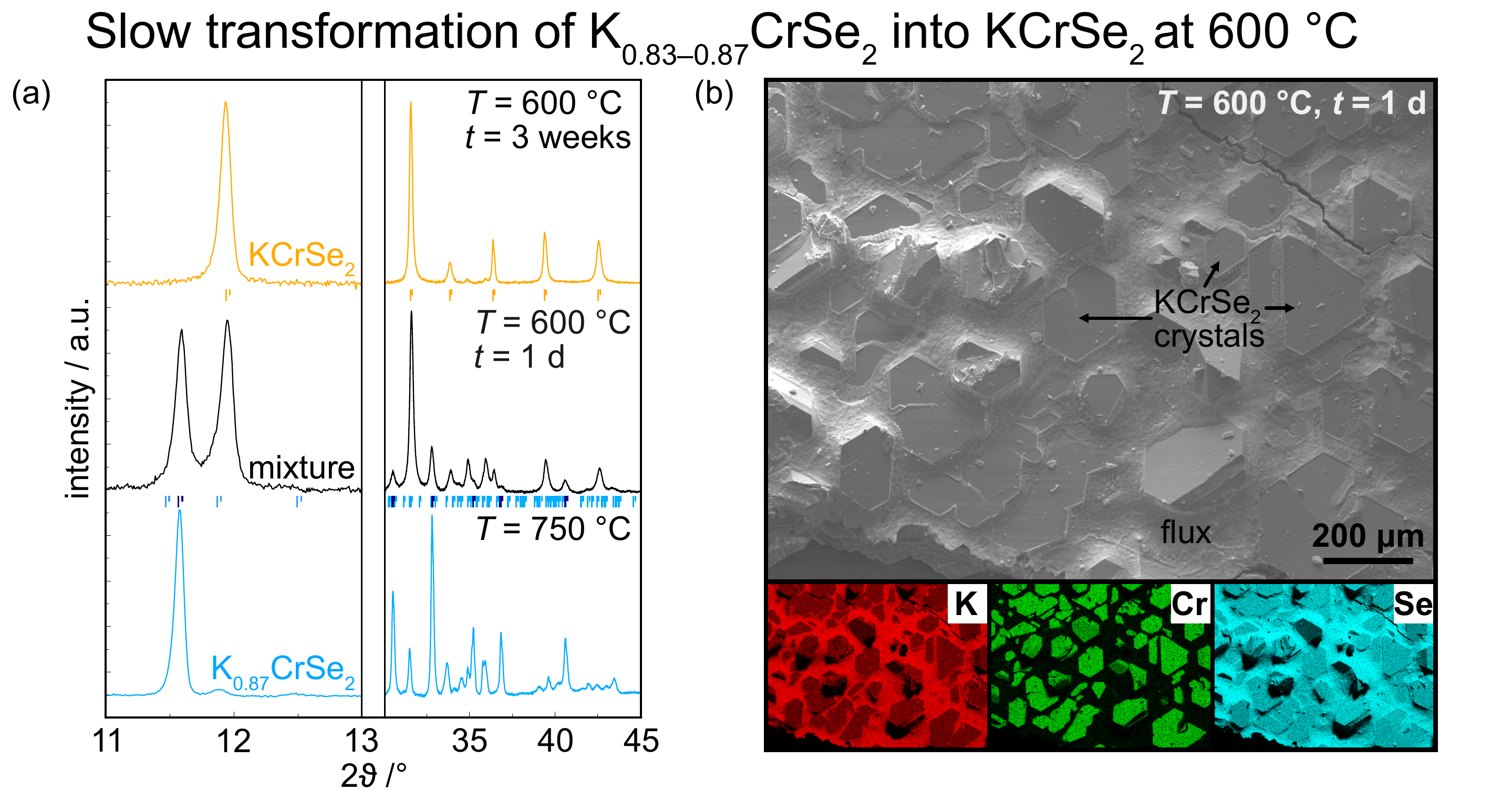}
\caption{The transformation of \ce{K_{0.83--0.87}CrSe2} to \ce{KCrSe2} (a) Selected areas of PXRD patterns of samples quenched from \textit{T} = 600 °C after \textit{t} = three weeks (top), \textit{T} = 600 °C, \textit{t} = 1 d (mid), and \textit{T} = 750 °C (bottom). The 2$\theta$ ranges on the x-axis have been chosen to highlight the multi-phase pattern in the middle. The complete diffraction patterns are displayed in the SI. Positions of \ce{KCrSe2}, \ce{K_{0.87}CrSe2} main and \ce{K_{0.87}CrSe2} satellite reflections are marked in orange, dark blue and blue, respectively. The background was subtracted for clarity. (b) SEM-SE picture taken at 20 kV acceleration voltage of sample quenched after less than one day at 600 °C. Recrystallizing shards of \ce{KCrSe2} are visible with a background of incompletely removed \ce{K2Se2} flux. EDS maps of the constituting elements are displayed at the bottom.}
\label{Fig3}
\end{center}
\end{figure*}

When synthesizing full-stoichiometric \ce{KCrSe2} from K/Se self-flux synthesis, we noticed that additional attention needs to be paid to the reaction time needed for the transformation of \ce{K_{0.83--0.87}CrSe2} to \ce{KCrSe2}. 
To illustrate this, we heated two synthetic batches of K, Cr, and Se (molar ratios 8:1:8) to 1000 °C and cooled them down to 600 °C at a rate of 3 °C/h. One batch was removed from the furnace and quenched after an annealing time of \textit{t} = 1 d at this temperature, while the second batch was annealed inside the flux for \textit{t} = 3 weeks before being quenched. 
In both the PXRD patterns (Fig. \ref{Fig3}(a)) and the magnetic properties of mm-sized crystal plates (see SI, Fig. S1), it can be seen clearly that after the short dwelling time at 600 °C, still significant amounts of incommensurately modulated \ce{K_{0.83--0.87}CrSe2} are present. After the elongated tempering, these disappear and only \ce{KCrSe2} remains as the reaction product. 
The newly transformed reaction product \ce{KCrSe2} hereby forms as hexagonal platelets on the surface of larger crystal plates (Fig. \ref{Fig3}(b)). This indicates that the under-stoichiometric phase is transformed by incorporation of K cations, available as excess in the flux. This apparently diffusion-limited process\cite{Song2021} takes extended amounts of time to fully transform the larger \ce{K_{1--\textit{x}}CrSe2} crystals, which had grown at higher temperatures. On the investigated surface, only \ce{KCrSe2} and residual \ce{K2Se2} flux were identified from EDS data, the \ce{K_{0.83--0.87}CrSe2} precursor is believed to remain at the center of the crystal piece.

\subsection{Direction-dependent magnetic properties of \ce{KCrSe2} single crystals}

\begin{figure*}
\begin{center}
\includegraphics[width=16.5cm]{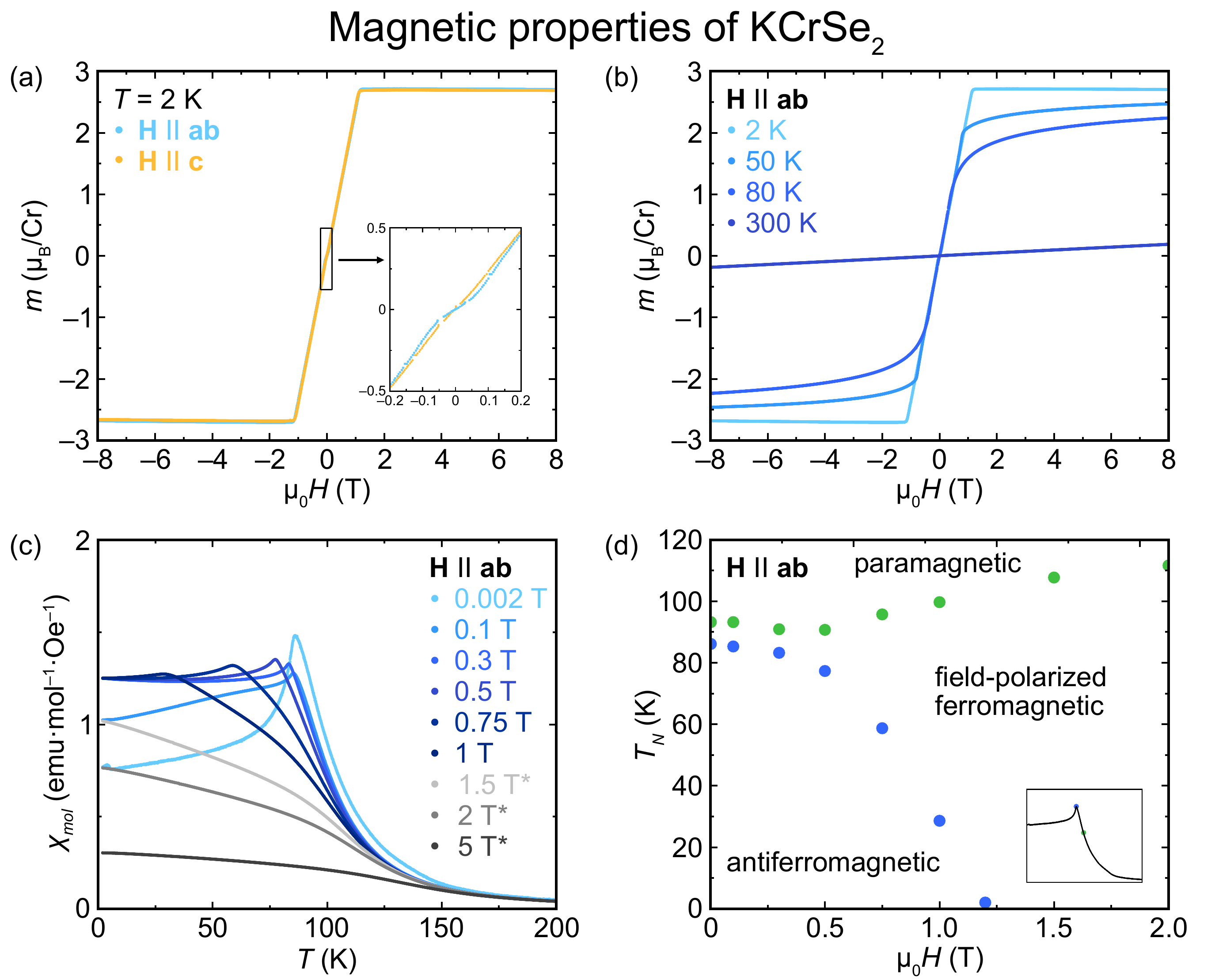}
\caption{Magnetic properties of oriented \ce{KCrSe2} single crystals. (a) Field-dependent magnetic moment at 2 K. The inset highlights the metamagnetic transition at very low fields for \textbf{H} $\parallel$ \textbf{ab}. Data were corrected for demagnetization effects. (b) Field-dependent magnetic moment at various temperatures for \textbf{H} $\parallel$ \textbf{ab}. (c) Temperature-dependent molar magnetic susceptibility at various external fields. (d) Field-dependence of the Néel temperature for \textbf{H} $\parallel$ \textbf{ab}.}
\label{Fig4}
\end{center}
\end{figure*}

We report the first direction-dependent magnetization measurements on single crystals of full-stoichiometric \ce{KCrSe2}, enabled by the extended annealing at 600 °C described above. Crystals were mounted with the magnetic field applied either parallel (\textbf{H} $\parallel$ \textbf{ab}) or perpendicular (\textbf{H} $\parallel$ \textbf{c}) to the layer plane.

In Fig. \ref{Fig4}(a), we show the field-dependent magnetization $m(H)$ at $T$ = 2 K, which displays a saturation at $\mu$\textsubscript{0} \textit{H}\textsubscript{sat}$\approx$ 1.2 T with a magnetic moment of 2.73 $\mu$\textsubscript{B}/Cr, consistent with the reported A-type AFM behavior. This saturation magnetic moment is lower than the theoretically expected value for a free \ce{Cr^{3+}} ion of $\mu$\textsubscript{eff} = 3.87 $\mu$\textsubscript{B}/Cr), but higher than the value for the incommensurately modulated \ce{K_{1–x}CrSe2}, where a saturation magnetization of 2.4 $\mu$\textsubscript{B}/Cr was found.\cite{Eder2025} The higher value of fully stoichiometric \ce{KCrSe2} is likely due to the lower average oxidation state and therefore higher number of available unpaired electrons of the Cr atoms. For \textbf{H} $\parallel$ \textbf{ab}, a subtle change in slope appears at very low fields, indicative of a spin-flop transition. 
This transition is absent for \textbf{H} $\parallel$ \textbf{c}, suggesting that the magnetic easy axis is oriented parallel to the layer plane. The shapes of the \textit{m}(\textit{H}) curves and their indentation due to the spin-flop are reminiscent of the delafossites \ce{NaCrS2} and \ce{AgCrSe2} with much larger fields involved for the reference phases.\cite{Huang2022, Baenitz2021} The closest relation concerning the saturation field to literature magnetization curves is found with \ce{NaCrSe2}, where for both 
(\textbf{H} $\parallel$ \textbf{ab}) and (\textbf{H} $\parallel$ \textbf{c}) a linear rise in magnetization until a saturation field at 3--3.5 T is reported.\cite{Huang2022}
For all three magnetically similar delafossites - \ce{NaCrS2}, \ce{AgCrSe2} and \ce{NaCrSe2} - the magnetic moments have been determined to lie within the layer plane by neutron powder diffraction.\cite{engelsman1973}
\textit{m}(\textit{H}) curves collected at higher temperatures (Fig. \ref{Fig4}(b)) show -- as expected -- a decreasing saturation moment. At room temperature (300 K), the paramagnetic behavior is evident. No hysteresis was observed in the $m(H)$ measurements at any temperature (Fig. \ref{Fig4}(a), Fig. S4).

Fig. \ref{Fig4}(c) shows the temperature-dependent magnetic susceptibility, calculated as $\chi$\textsubscript{mol} = $\frac{m}{H*n}$. The sharp peak marks the transition from a paramagnetic to an antiferromagnetic state at the N\'eel temperature \textit{T}\textsubscript{N} = 86 K (at $\mu_0 H$ = 2 mT). This value aligns with recent reports \cite{Song2021} but contrasts with earlier data suggesting a significantly lower \textit{T}\textsubscript{N} of 40 K \cite{Fang_1996}. As shown in Fig. \ref{Fig4}(d), the peak in the magnetic susceptibility decreases rapidly with increasing external field, reaching approximately 60 K at $\mu_0H = 0.75$ T. This behavior resolves the apparent discrepancy in earlier reports and can be correlated with the remarkably low fields required for the metamagnetic transition in this material. This is in agreement with the earlier analysis of \textit{Song et al.}\cite{Song2021} At external fields higher than the saturation field of $\sim$1.2 T, the $\chi_{mol}(T)$ curves exhibit a form of plateauing at low T compared to the high-temperature paramagnetic behavior. This change corresponds to the transition from the paramagnetic to the field-polarized state of ferromagnetic spin alignments. 
We have extracted this transition as a minimum of the first derivative of $\frac{d\chi_{mol}}{dT}$ and included it in Fig. \ref{Fig4}(d).  

While metamagnetic transitions are common in many delafossite-like antiferromagnets due to their pronounced 2D nature, the fields needed to induce a metamagnetic transition in \ce{KCrSe2} are remarkably low. As comparison, for \ce{NaCrS2}, the spin-flop transition finishes at fields of  $\sim$3 T \cite{Huang2022},  and for \ce{AgCrSe2}, around 6--7 T \cite{Baenitz2021}. However, the saturation fields of both phases are significantly higher than in \ce{KCrSe2} as well. This highlights the unusually soft AFM ground state of \ce{KCrSe2} and suggests a close energetic proximity to the ferromagnetic ground state configuration.

The Curie-Weiss temperature $\theta_{CW}$ was determined from fitting the reciprocal susceptibility (Fig. S4(d)) in the 200--299 K temperature range and amounts to approximately 140 K, which is similar to the most recent investigation \cite{Song2021}. A slight under-estimation could be due to the fact that no perfect linear behavior of $1/\chi$(T) is established in this temperature range yet. The combination of the positive $\theta_{CW}$, indicating ferromagnetic nearest-neighbor interactions, and the AFM behavior is in line with the A-type AFM magnetic ordering, which has been established by previous investigations. A similar situation was observed in the modulated \ce{K_{0.87}CrSe2}, which orders antiferromagnetically as well with a higher $T_N$ (135 K) and $\theta_{CW}$ (180 K), but a much stronger anisotropy in the magnetic responses.\cite{Eder2025}

Previous DFT-calculations have indicated that the AFM ordered state in \ce{KCrSe2} is energetically only marginally lower than an FM state with energy differences within the standard deviation of their calculations.\cite{Fang_1996} This would align with the fact that \ce{KCrSe2} has a higher magnetic ordering temperature than most related \textit{A}Cr\textit{X}\textsubscript{2} compounds, but a much lower saturation field. We believe that the observed facile suppression of AFM order or rather transformation to an FM region by the external field is a result of this energetically very similar magnetically ordered states.

\section{Conclusion}

We have demonstrated that subtle changes in quenching temperature during self-flux synthesis can reproducibly tune the formation of three distinct \ce{K_{1-–\textit{x}}CrSe2} phases: stoichiometric \ce{KCrSe2}, the incommensurately modulated \ce{K_{0.87}CrSe2}, and vacancy-disordered \ce{K_{0.68--0.65}CrSe2}. This finding resolves inconsistencies in the literature by linking structural and compositional variations to specific synthetic conditions in the K--Cr--Se system.

Our results show that all three phases can be accessed from the same nominal composition (K:Cr:Se = 8:1:8), with quenching temperature acting as the primary control parameter. For the most potassium-poor phase \ce{K_{0.65--0.68}CrSe2}, a reduction of the K-content of the flux to (K:Cr:Se = 5:1:8) was necessary to obtain phase-purity.
All three phases could be obtained as high-quality single crystals, enabling the first SXRD refinement of \ce{K_{0.68(2)}CrSe2} and \ce{K_{0.65(3)}CrSe2}. For stoichiometric \ce{KCrSe2}, we could furthermore perform the first direction-dependent magnetization measurements, revealing a low-field spin-flop transition and a highly field-sensitive Néel temperature -- signatures of a near-degeneracy between AFM and FM ground states, which explains earlier reports on differing transition temperatures.

Our findings highlight the potential of flux synthesis as a tool not just for crystal growth but for rational phase control and discovery in layered materials. We find that understanding and controlling intermediate phases is key to navigating complex phase spaces and to tuning the structural and magnetic properties of delafossite-type systems.

\begin{acknowledgement}

This work was supported  by the Swiss National Science Foundation under Grants No. PCEFP2\_194183 and No. 200021-204065. The help of Stefano Gariglio in the setup and performance of the PXRD measurements is thankfully acknowledged.

\end{acknowledgement}

\begin{suppinfo}

Additional details on determination of elemental composition, further PXRD patterns and complementary magnetic data.

\end{suppinfo}

\bibliography{KCrSe2}

\end{document}